\documentclass[prl,twocolumn,showpacs]{revtex4-1}
\usepackage[utf8x]{inputenc}
\usepackage{amsbsy}
\usepackage{graphicx}
\usepackage{color}
\usepackage[dvipsnames]{xcolor}
\usepackage{dcolumn}
\usepackage{amsmath}
\usepackage{amssymb}
\usepackage{amsfonts}
\usepackage{subfigure}

\usepackage[colorlinks=true, citecolor=black, linkcolor=black, urlcolor=black]{hyperref}

\newcommand{\beq}{\begin{equation}}
\newcommand{\eeq}{\end{equation}}
\newcommand{\beqn}{\begin{eqnarray}}
\newcommand{\eeqn}{\end{eqnarray}}

\newcommand{\tcdw}{$T_{\mathrm{CDW}}$}
\newcommand{\qcdw}{$\mathbf{q}_{\mathrm{CDW}}$}
\newcommand{\rscha}{\boldsymbol{\mathcal{R}}}
\newcommand{\phischa}{\boldsymbol{\Phi}}

\newcommand{\Bigtr}[1]{\textup{Tr}\Bigl[\,#1\,\Bigr]}

\newcommand{\Avg}[1]{\left\langle #1\right\rangle_{\rhoscha}}
\newcommand{\rhoscha}{\rho_{\rscha,\phischa}}
\newcommand{\sss}[1]{\scriptscriptstyle{\text{#1}}}
\newcommand{\bR}{\boldsymbol{R}}
\newcommand{\Eel}{E_{\sss{el}}}
\newcommand{\Fel}{F_{\sss{el}}}
\newcommand{\Sel}{S_{\sss{el}}}
\newcommand{\Sion}{S_{\sss{ion}}}

\begin{document}
\title{Weak Dimensionality Dependence and Dominant Role of Ionic Fluctuations in the Charge-Density-Wave Transition of NbSe$_2$}

\author{Raffaello Bianco$^{1}$}
\email{raffaello.bianco@ehu.eus}
\author{Lorenzo Monacelli$^{2,3}$}
\author{Matteo Calandra$^{3,4,5}$}
\author{Francesco Mauri$^{2,3}$}
\author{Ion Errea$^{1,6,7}$}
\email{ion.errea@ehu.eus}

\affiliation{$^1$Centro de F\'isica de Materiales (CSIC-UPV/EHU), Manuel de Lardizabal pasealekua 5, 20018 Donostia/San Sebasti\'an, Spain}
\affiliation{$^2$Dipartimento di Fisica, Università di Roma Sapienza, Piazzale Aldo Moro 5, I-00185 Roma, Italy}
\affiliation{$^3$Graphene Labs, Fondazione Istituto Italiano di Tecnologia, Via Morego, I-16163 Genova, Italy}
\affiliation{$^4$ Dipartimento di Fisica, Università di Trento, Via Sommarive 14, 38123 Povo, Italy.}
\affiliation{$^5$ Sorbonne Universit\'e, CNRS, Institut des Nanosciences de Paris, UMR7588, F-75252 Paris, France}
\affiliation{$^6$Fisika Aplikatua 1 Saila, Gipuzkoako Ingeniaritza Eskola, University of the Basque Country (UPV/EHU),
Europa Plaza 1, 20018 Donostia/San Sebasti\'an, Spain}
\affiliation{$^7$Donostia International Physics Center (DIPC), Manuel Lardizabal pasealekua 4, 20018 Donostia/San Sebasti\'an, Spain}

\begin{abstract}
Contradictory experiments have been reported about the dimensionality effect on the charge-density-wave transition in 2H NbSe$_2$. While scanning tunnelling experiments on single layers grown by molecular beam epitaxy measure a charge-density-wave transition temperature in the monolayer similar to the bulk, around 33 K, Raman experiments on exfoliated samples observe a large enhancement of the transition temperature up to 145 K. By employing a non-perturbative approach to deal with anharmonicity, we calculate from first principles the temperature dependence of the phonon spectra both for bulk and monolayer. In both cases, the charge-density-wave transition temperature is estimated as the temperature at which the phonon energy of the mode driving the structural instability vanishes. The obtained transition temperature in the bulk is around 59 K, in rather good agreement with experiments, and it is just slightly increased in the single-layer limit to 73 K, showing the weak dependence of the transition on dimensionality. Environmental factors could motivate the disagreement between the transition temperatures reported by experiments. Our analysis also demonstrates the predominance of ionic fluctuations over electronic ones in the melting of the charge-density-wave order.
\end{abstract}

%\pacs{74.25.Kc,74.90.+n,67.63.-r,67.80.F-,63.20.Ry,31.15.A-}

\maketitle

A charge-density wave (CDW) is a structural distortion of the crystal lattice that induces a modulation of the electronic density~\cite{RevModPhys.60.1129}. CDWs in transition metal dichalcogenides (TMDs)~\cite{doi:10.1080/00018736900101307} are particularly relevant because they seem to compete with superconductivity~\cite{doi:10.1080/00018737500101391,Sipos2008,Smith_1972,PhysRevLett.95.117006,PhysRevB.87.134502,PhysRevLett.103.236401,Morosan2006} and form a phase diagram similar to the high-temperature superconductors~\cite{Paglione2010,Varma2010}. The origin of the CDW instability and its interaction with superconductivity in TMDs continues to be a topic on intense debate~\cite{PhysRevLett.106.196406,PhysRevB.92.140303,Kogar1314,PhysRevB.100.205423}, as it might enlighten the hidden physics in strongly correlated materials where different phases compete.   

\begin{figure}[t]
\includegraphics[width=1.0\columnwidth]{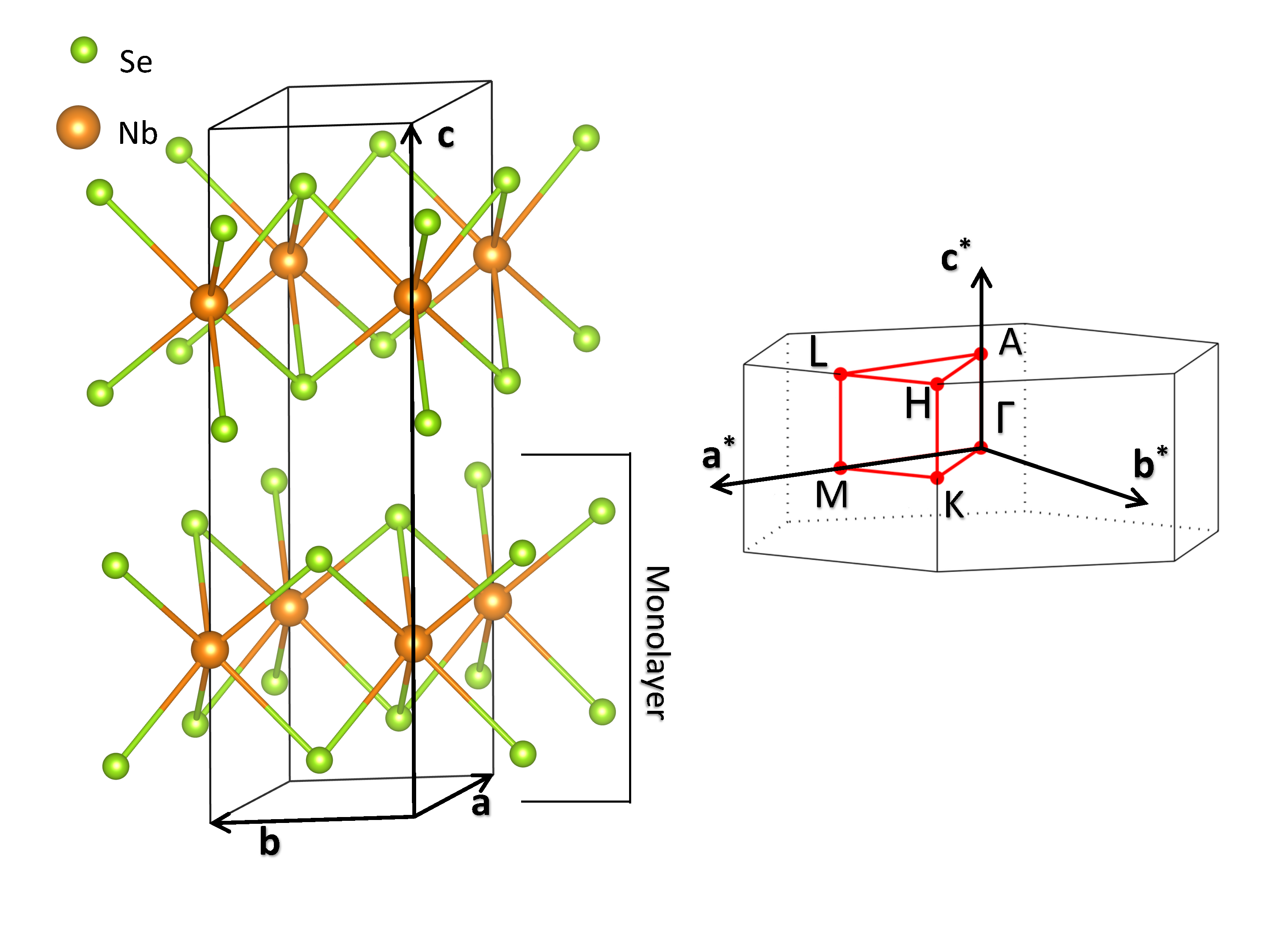}
\caption{Crystal structure of bulk NbSe$_2$ in the 2H polytype. The unit cell is depicted, which is formed by two non-equivalent layers. The crystal structure of the monolayer is marked in the figure, which only contains one single layer. The lattice paremeters used are given in the Supplementary Material~\cite{supp}. In the right panel the Brillouin zone is shown, which is restricted to the  $\Gamma$MK plane in the monolayer.} 
\label{fig1}
\end{figure}

TMDs tend to adopt layered crystal structures. Each layer is formed by transition metal atoms sandwiched by covalently bonded chalcogens (see Fig. \ref{fig1}). Due to the weak van der Waals interaction that holds together the layers in the bulk, TMDs can be exfoliated down to the monolayer~\cite{Novoselov26072005}. Monolayer TMDs can currently also be synthesized by chemical means~\cite{Manzeli2017}. This has opened the possibility to study the effect of dimensionality on CDW transitions. The results obtained thus far, however, do not show a clear trend and it is generally not clear whether experimental results are affected by environmental factors. In monolayer TaS$_2$ the CDW present in the bulk disappears~\cite{PhysRevB.98.035203}, while in the isolectronic and isostructural TaSe$_2$ it remains unchanged~\cite{doi:10.1021/acs.nanolett.7b03264}. Even if NbS$_2$ has no CDW transition in the bulk~\cite{PhysRevB.86.155125}, a CDW transition has been observed in the monolayer grown on top of bilayer graphene~\cite{Lin2018}. This result seems substrate dependent, as a monolayer grown on Au(111) does not show any CDW feature down to 30 K~\cite{PhysRevMaterials.3.044003}. Similarly, in monolayer TiSe$_2$ the CDW temperature (\tcdw) is enhanced with respect to the bulk, but the value of \tcdw\ is strongly substrate dependent~\cite{Chen2015,Kolekar_2017}. 

The effect of thickness on the \tcdw\ in NbSe$_2$ is even more controversial. The most stable polytype of NbSe$_2$ is the 2H shown in Fig. \ref{fig1}, which undergoes a CDW transition at 33 K to an incommensurate structure very close to a 3$\times$3$\times$1 ordering~\cite{doi:10.1080/00018737500101391,ja3120554}. Inelastic x-ray experiments evidence that the CDW transition is second order. In fact, a longitudinal acoustic (LA) phonon mode collapses exactly at 33 K at $\mathbf{q}_{\mathrm{CDW}} \sim 2/3\,\Gamma M$~\cite{PhysRevLett.107.107403}, a momentum consistent with the periodicity of the CDW phase. In other TMDs as well, the phonon frequency of the mode that drives the CDW is strongly reduced by cooling and eventually vanishes at \tcdw~\cite{PhysRevLett.86.3799,PhysRevLett.107.266401}. The CDW transition temperature in the NbSe$_2$ monolayer has been determined by Raman measurements in exfoliated samples on sapphire substrates~\cite{Xi2015,lin2020patterns}, as well as by scanning tunnelling microscopy (STM) in single-layers grown by molecular beam epitaxy (MBE) on bilayer graphene, which confirmed that the CDW order remains 3$\times$3~\cite{Ugeda2016}. The problem is that while the Raman experiments on exfoliated samples estimate a huge enhancement of \tcdw\ up to 145 K~\cite{Xi2015,lin2020patterns}, STM experiments determine that dimensionality does not affect \tcdw\ as the CDW occurs between 25~K and 45~K~\cite{Ugeda2016}. 

In this work we present first-principles calculations of \tcdw\ both in bulk and monolayer NbSe$_2$. We determine that the intrinsic CDW transition temperature is barely affected by dimensionality. The theoretically calculated phonon spectrum and \tcdw\ in the bulk are in good agreement with inelastic x-ray experiments. Since the value obtained for the monolayer in a completely comparable calculation is very similar, it is confirmed that bulk and suspended monolayer NbSe$_2$ are expected to have a similar CDW transition temperature as suggested by the STM experiments. Our study also demonstrates that, when anharmonicity is fully taken into account, the contribution to the melting of the CDW order given by the electronic thermal fluctuations is totally irrelevant compared to the contribution of the ionic thermal fluctuations. 

\begin{figure}[t]
\includegraphics[width=1.0\columnwidth]{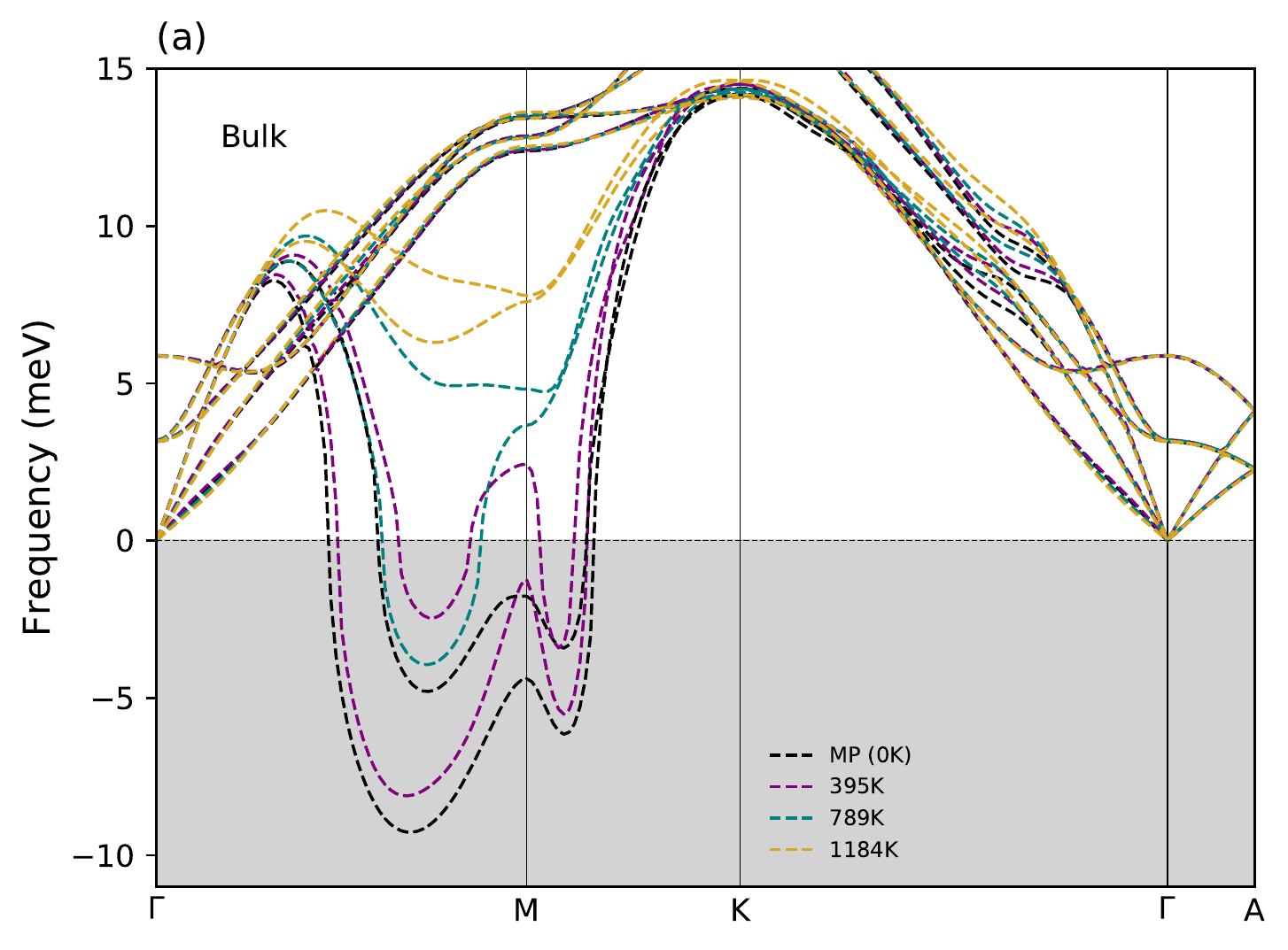} \\
\includegraphics[width=1.0\columnwidth]{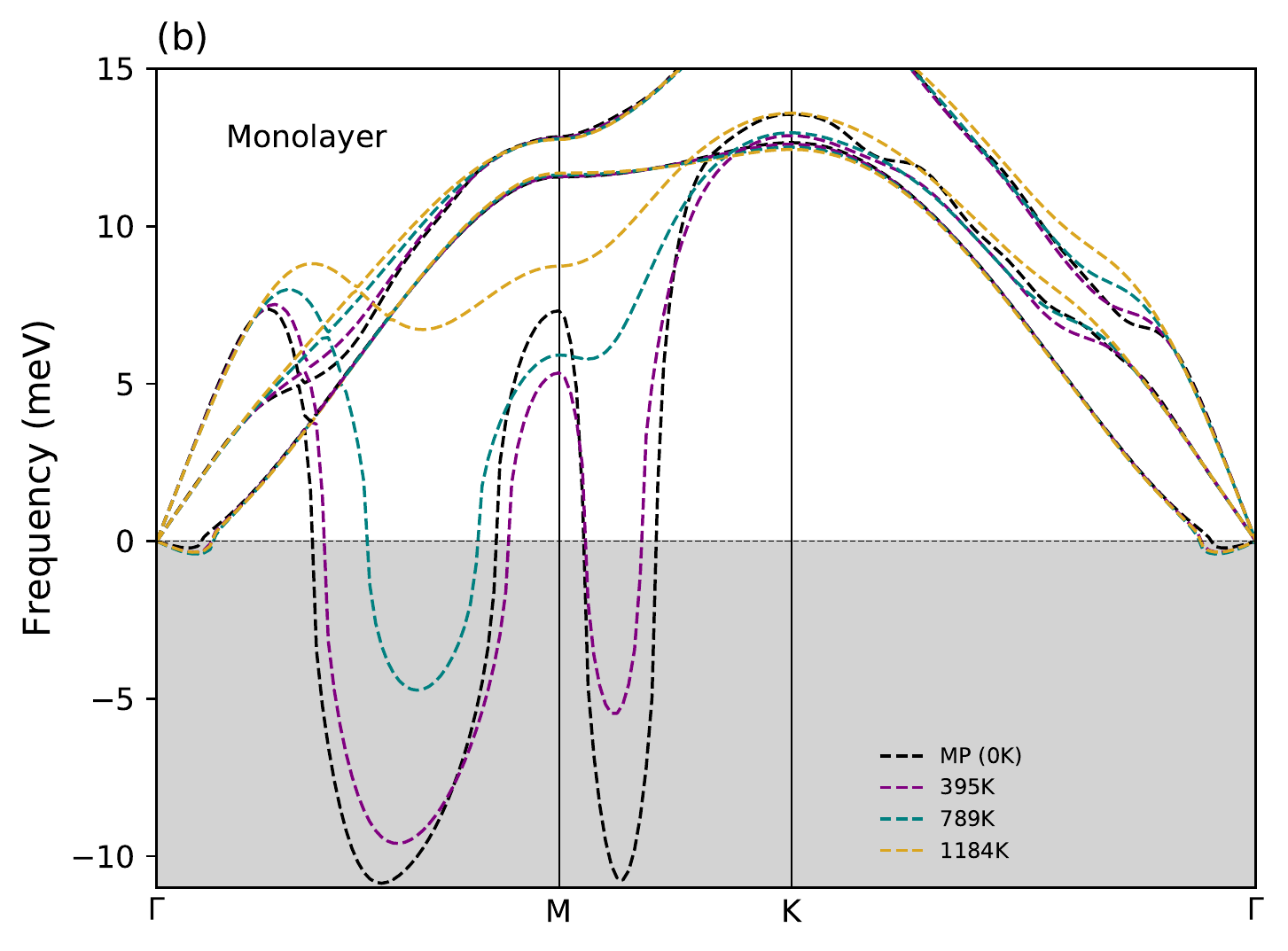}
\caption{Harmonic phonon spectra at several temperatures for (a) bulk and (b) monolayer NbSe$_2$.  
The $0$~K dispersions are obtained with the Methfessel and Paxton (MP) cold smearing.  
The finite-temperature results are estimated using the Fermi-Dirac occupation of the Khon-Sham states, 
within the grand-canonical extension of DFT (see main text and Supplementary Material~\cite{supp}). 
The grey areas represent imaginary phonon frequencies, which are given with negative values.
%\edit{Notice the strong similarity between the bulk and monolayer imaginary phonon dispersion at 789~K}.
} 
\label{fig2}
\end{figure}

In Fig.~\ref{fig2} we show the (0~K) harmonic phonon spectra calculated for the bulk and the monolayer (obtained with the Methfessel and Paxton~\cite{PhysRevB.40.3616} cold smearing technique, see Supplementary Material~\cite{supp}). As it has been already pointed out~\cite{PhysRevB.80.241108,doi:10.1021/acs.nanolett.8b00237,PhysRevLett.107.107403}, in both cases the harmonic phonons show many unstable modes. Following the displacement pattern of any of them, the ionic potential energy surface $V(\bR)$ is lowered. Even if in both cases the LA mode is unstable close to \qcdw$=2/3\,\Gamma M$, there are some differences in the harmonic phonon spectra. The most unstable mode along $\Gamma M$ in the monolayer is shifted to smaller momentum with respect to the bulk, around $0.56\,\Gamma M$, in agreement with previous calculations~\cite{PhysRevB.80.241108}. Remarkably, there is no instability at the $M$ point in the monolayer, but a deep instability emerges along $MK$. Even if it has been argued that the CDW spatial modulation can be inferred from the $\mathbf{q}$ point where the deepest instability occurs in the (0~K) harmonic calculation~\cite{PhysRevB.80.241108,doi:10.1021/acs.nanolett.8b00237}, this can only be understood as a first hint and it may yield to a wrong interpretation. As a matter of fact, the modulation of the CDW can only be determined theoretically by calculating the phonon spectra as a function of temperature and seeing at which $\mathbf{q}$ a phonon becomes unstable on cooling.

Calculating temperature-dependent phonon frequencies in systems that undergo a second-order structural phase transition upon cooling is not a trivial task. A first, elementary, approach is to perform a temperature-dependent analysis within the harmonic approximation~\cite{PhysRevLett.121.226602,doi:10.1021/acsnano.6b07737,PhysRevB.92.245131}. We perform these calculations within Mermin's grand-canonical extension to density-functional theory (DFT)~\cite{PhysRev.137.A1441}. We use the Fermi-Dirac distribution function to obtain the finite-temperature occupation of the Khon-Sham states~\cite{PhysRevB.51.6773}, and we adopt the natural approximation of using the ground-state exchange-correlation functional also at finite temperature~\cite{PhysRevLett.101.070401,PhysRevLett.105.085501}. In this way, the finite-temperature electronic free energy $\Fel(\bR)=\Eel(\bR)-T\Sel(\bR)$, as a function of the ionic configuration $\bR$, is obtained (with $\Eel(\bR)$ we denote the average electronic energy at the considered temperature, which coincides with the electronic ground state energy $V(\bR)$ in the zero-temperature limit). The Hessian (divided by the square root of the masses) of the electronic free energy around its minimum gives the finite-temperature generalization of the standard 0~K harmonic dynamical matrix (for which the Hessian of $V(\bR)$ around its minimum is considered). An imaginary phonon thus corresponds to an atomic displacement pattern that lowers the electronic free energy and drives a displacive second-order phase transition. 

It is worthwhile to stress that, at this level, thermal (and quantum) fluctuations of ions are not taken into account, thus only thermal (and quantum) fluctuations of electrons can play a role in the melting of the CDW order. As shown in Figs.~\ref{fig2} and~\ref{fig4}, the finite-temperature electronic contribution to the free energy melts the CDW at around 872~K for the bulk and at around 917~K for the monolayer, which are both very far from the \tcdw's reported experimentally. On the other hand, it is interesting to observe that the differences in the $\mathbf{q}$-order of the instabilities between bulk and monolayer found in the harmonic results at 0~K become irrelevant: the temperature-dependent harmonic analysis shows that, in agreement with experiments, in both bulk and monolayer the instability appears close to \qcdw$=2/3\,\Gamma M$. 

The harmonic results show a very weak dimensionality dependence in both \qcdw\ and \tcdw\ (the monolayer and bulk temperature differ by 5\%). However, while the ordering vector \qcdw\ is in good agreement with experiments, the poor agreement of the calculated \tcdw's makes very difficult to draw any reliable conclusion at the harmonic level, and suggests that anharmonicity could be the key ingredient in the CDW melting, as it happens in other chalcogenides~\cite{PhysRevB.92.140303,doi:10.1021/acs.nanolett.9b00504,doi:10.1021/acs.nanolett.0c00597,PhysRevB.97.014306,PhysRevLett.122.075901,PhysRevB.100.214307}. Since the high-temperature undistorted phase is not a minimum of the Born-Oppenheimer potential $V(\bR)$, anharmonicity cannot be included through standard perturbative approaches on top of the harmonic result. We overcome this problem by making use of the stochastic self-consistent harmonic approximation (SSCHA)~\cite{PhysRevB.89.064302,PhysRevB.96.014111,PhysRevB.98.024106}, a quantum free energy variational method that allows to perform full anharmonic non-perturbative calculations. Within the Born-Oppenheimer approximation, if the electronic temperature fluctuations are neglected, the nuclei Hamiltonian is given by $K+V(\bR)$, where $K$ is the kinetic 
energy operator. With the SSCHA we evaluate the free energy $F=E-TS$ of the system by minimizing the free energy functional 
\begin{equation}
F[\rhoscha]=\Avg{K+V(\bR)}-T\,\Sion\left[\rhoscha\right]\,,
\end{equation}
where $\rhoscha$ is the ionic density matrix of an auxiliary harmonic potential parametrized with average ionic positions $\rscha$ and effective force constants $\phischa$ (related to the amplitude of the ionic fluctuations around~$\rscha$), 
$\Avg{\Box}=\Bigtr{\rhoscha\,\Box}$, and $\Sion[\rhoscha]=-\Avg{\textup{ln}\,\rhoscha}$ is the ionic entropy. Therefore, at this level, the system's entropy $S$ includes only the ionic contribution $\Sion$.
% 
% without approximating $V$ with respect to centroid positions $\rscha$, which determine the average ionic positions, and effective force constants $\phischa$, which are related to the amplitude of the ionic fluctuations around $\rscha$. 
%\edit{As for the harmonic case, since we are only interested in characterizing the phase transition, the anharmonic phonon frequencies are obtained by diagonalizing the Hessian of free energy~\cite{PhysRevB.96.014111}.} 
%The calculation of phonons following this approach has already been successfully applied to TMDs~\cite{PhysRevB.92.140303,doi:10.1021/acs.nanolett.9b00504,doi:10.1021/acs.nanolett.0c00597}, 
%where phonon frequencies in agreement with experiments have been obtained, and to other calchogenides~\cite{PhysRevB.97.014306,PhysRevLett.122.075901}. 
The application of the SSCHA requires the calculation of atomic forces on supercells. We used 6$\times$6$\times$1 supercells for both the bulk and the monolayer, which are commensurate with \qcdw$=2/3\,\Gamma M$, and the forces were calculated on these supercells using DFT (see Supplementary Material\cite{supp} for further details on the calculations). 

\begin{figure}[t]
\includegraphics[width=1.0\columnwidth]{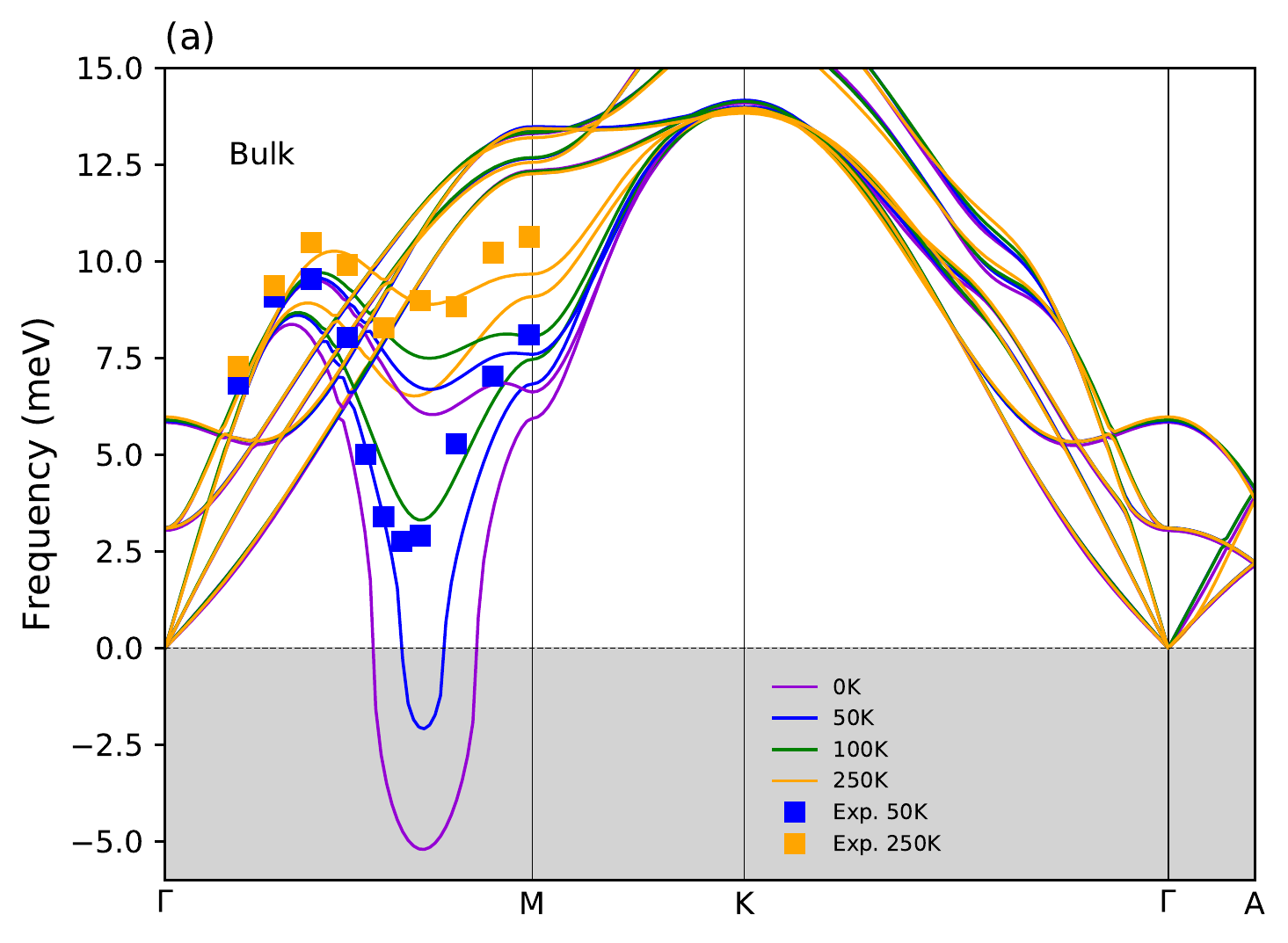} \\
\includegraphics[width=1.0\columnwidth]{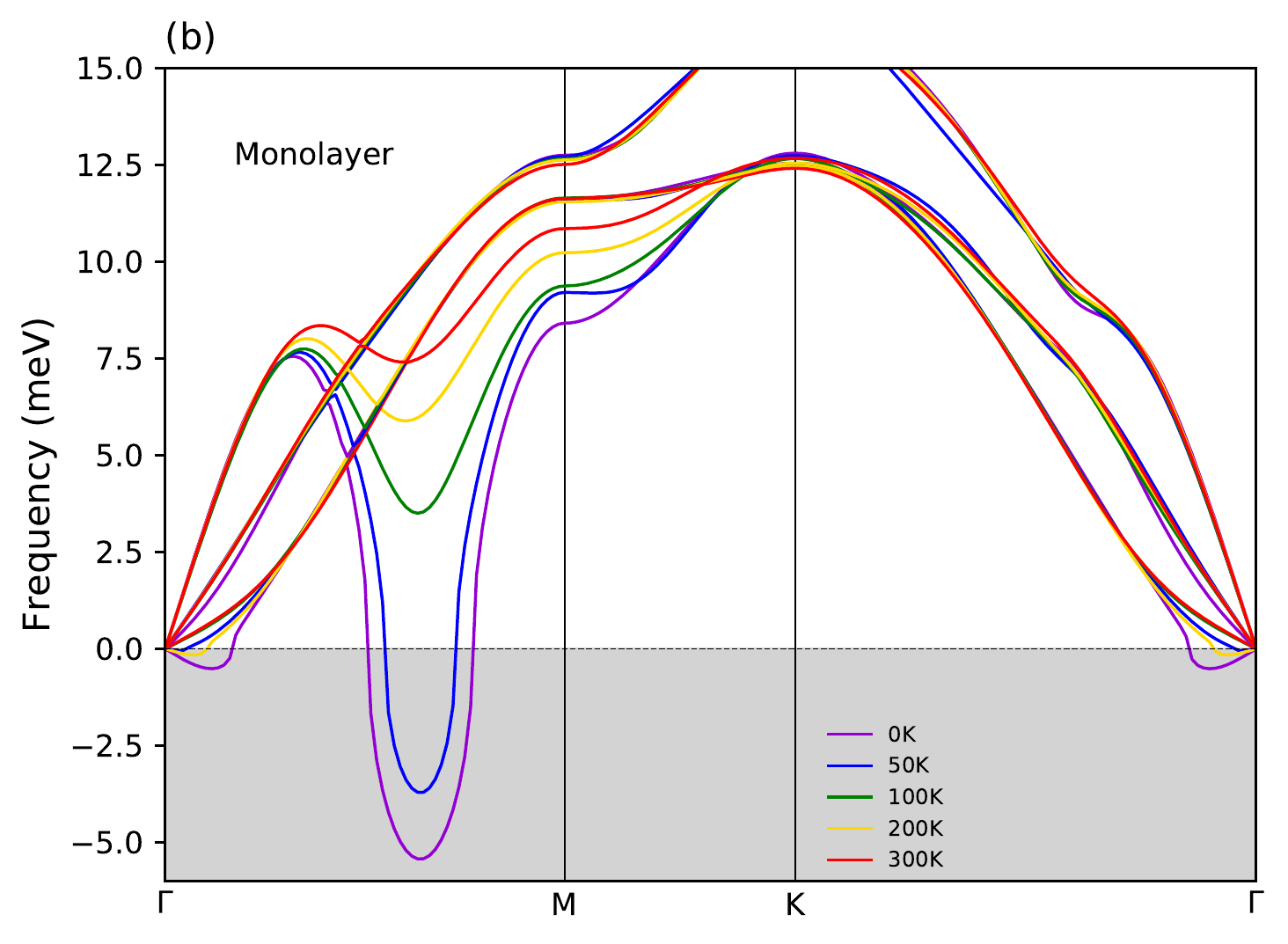}
\caption{Anharmonic phonon spectra calculated from the SSCHA free energy Hessian at several temperatures for (a) bulk and (b) monolayer NbSe$_2$. The grey areas denote imaginary phonon frequencies. In the bulk case the results are compared with the experimental values obtained with inelastic x-ray scattering~\cite{PhysRevLett.107.107403}.} 
\label{fig3}
\end{figure}

\begin{figure}[t]
\includegraphics[width=1.0\columnwidth]{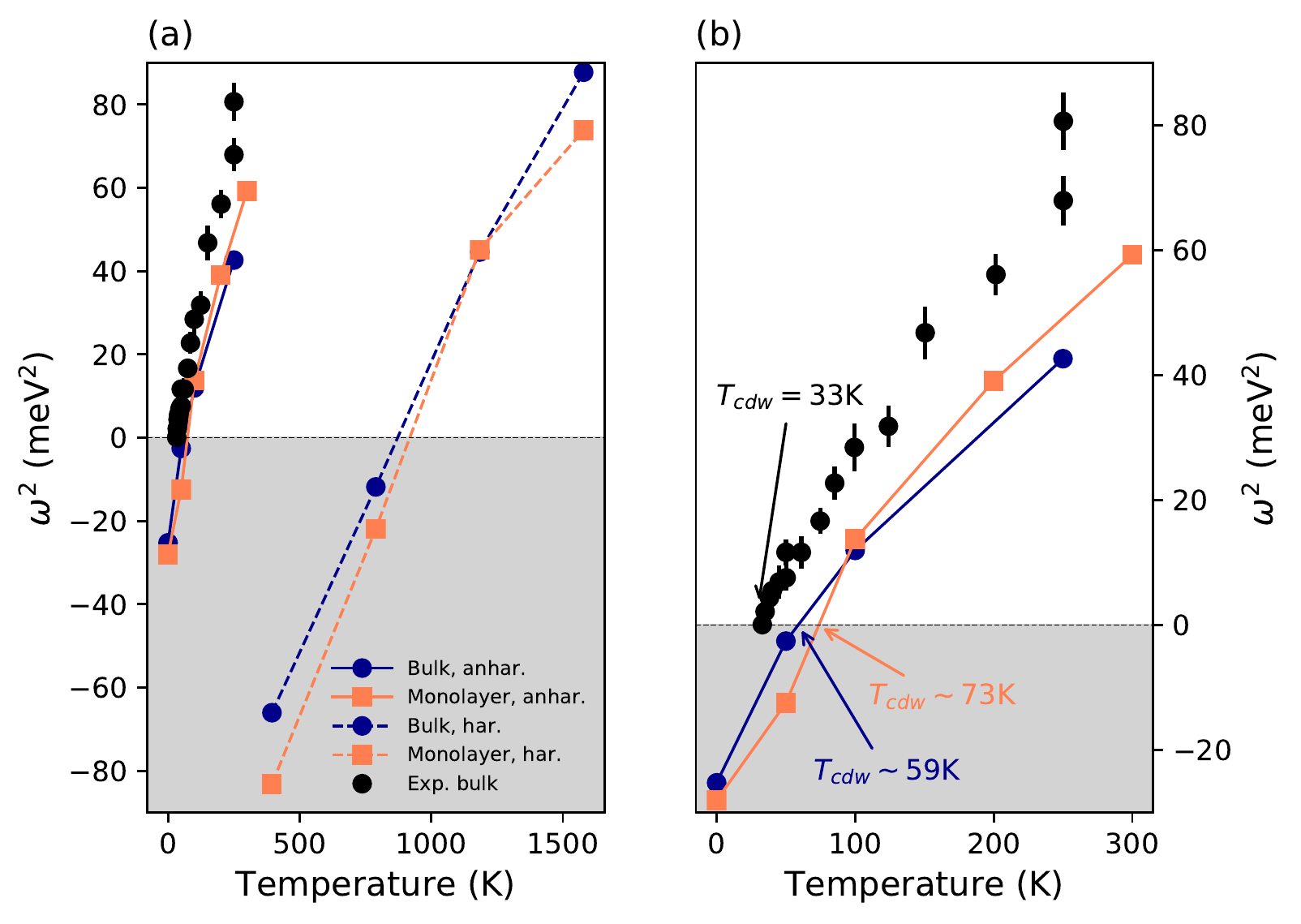} 
\caption{(a) Squared harmonic (dashed lines) and anharmonic (SSCHA, solid lines) frequencies of the longitudinal acoustic mode at \qcdw$=2/3\,\Gamma M$ as a function of temperature, obtained from the free energy Hessian. The harmonic results take into account only electronic fluctuations and do not capture at all the experimental trend~\cite{PhysRevLett.107.107403}. The anharmonic results include the ionic contribution too (the electronic contribution appears to be negligible though, see main text), and are considerably closer to the experimental results. (b) Zoom in on the anharmonic results. The \tcdw\ is estimated at 59 K for the bulk and at 73 K for the monolayer. The experimental CDW occurs at 33 K~\cite{PhysRevLett.107.107403}. The grey area denotes imaginary phonon frequencies in both figures.} 
\label{fig4}
\end{figure}

In Fig~\ref{fig3} we show the phonon spectra along a path, computed by Fourier interpolating the SSCHA free energy Hessian obtained from the  6$\times$6$\times$1 supercell
calculations. The anharmonic spectra display a huge anharmonic renormalization for the low-energy modes, with a remarkable temperature dependence. This highlights the relevant role of ionic fluctuations, which strongly feel the anharmonic part of the potential, in the CDW order melting. In the bulk case, the obtained phonon spectra at 250 K is in very good agreement with inelastic x-ray experiments~\cite{PhysRevLett.107.107403}. At 50 K our calculations are in rather good agreement with experiments as well, but underestimate the frequency of the LA mode at \qcdw$=2/3\Gamma M$, which it is still imaginary. The LA mode stabilizes between 50 K and 100 K in our calculations. Similarly, this mode also stabilizes in a similar temperature range in the monolayer. Interestingly, the deep instability along $MK$ in the monolayer is washed out by anharmonicity at all temperatures.

Now we focus our attention on the determination of the CDW modulation and transition temperature in this framework. The anharmonic phonon spectra in Fig. \ref{fig3} are obtained by Fourier interpolating the SSCHA free energy Hessian obtained in the 6$\times$6$\times$1 supercell. In this Fourier interpolated spectra, the CDW instability occurs close to \qcdw$=2/3\,\Gamma M$ in both bulk and monolayer, thus driving a 3$\times$3 CDW ordering in agreement with experiments. However, calculations in a larger 9$\times$9$\times$1 supercell for the monolayer show that the Fourier interpolated spectra are not converged with the supercell size over the whole Brillouin zone (see Supplementary Material~\cite{supp}). As a consequence, inferring the CDW ordering vector from the sole Fourier interpolated spectra would be risky. Nevertheless, the 3$\times$3 ordering of the CDW is an experimental fact in both bulk and monolayer, consistent as well with the temperature-dependent harmonic calculation, where we are not limited to a coarse interpolation grid. Moreover, the phonon frequency of the LA mode at \qcdw$=2/3\,\Gamma M$ is well converged with the supercell size (as confirmed in the monolayer case by the calculations on the 9$\times$9$\times$1 supercell~\cite{supp}). For this reason, we can confidently estimate \tcdw\ within SSCHA, and readily compare it between the bulk and the monolayer, by studying the temperature dependence of the obtained anharmonic phonon frequency of the LA mode at \qcdw$=2/3\,\Gamma M$.

As shown in Fig.~\ref{fig4}, the square of the calculated anharmonic phonon frequency of the LA mode at \qcdw$=2/3\,\Gamma M$ shows a temperature dependence in agreement with the experimental trend~\cite{PhysRevLett.107.107403}. The frequencies are slightly underestimated and, consequently, the value of the theoretical \tcdw\ is around $\sim$59 K, close to the experimental value of 33 K, but slightly overestimated due to the presence of systematic temperature-independent DFT-related errors. In the monolayer the frequency of the LA mode at \qcdw, as well as its temperature dependence, is practically on top of the bulk result. The CDW temperature in the monolayer is consequently very close to the bulk result, $\sim$73 K. Considering that the SSCHA calculations in the bulk and in the monolayer are performed with the same supercell, with consistent DFT parameters, the comparison between the results is perfectly justified.  We can thus conclude that the CDW temperature in NbSe$_2$ is weakly dependent on the dimensionality, supporting the results obtained with STM experiments~\cite{Ugeda2016}. 

The weak dimensionality dependence of the CDW in NbSe$_2$ might suggest a weak interlayer interaction. However, the large energy difference between the two softened phonon modes in the bulk at \qcdw, which are imaginary in the 0K harmonic calculations and have a similar but not identical distortion pattern for each layer (see Supplementary Material~\cite{supp}), shows that there is a non-negligible interlayer interaction. Therefore, the electronic screening, the electron-phonon coupling, and the electronic/ionic fluctuations are expected to play a different role in monolayer and bulk, and it is the complicate interplay between these effects that yields a very similar CDW phonon branch regardless of the thickness.

A CDW phase, as any order in condensed matter, melts with increasing temperature due to entropy or, in other words, fluctuations. 
The presented SSCHA calculations included the entropic effects coming from the ionic 
fluctuations only, with the electrons kept at 0~K.
The good agreement between the experiments and the \tcdw\ obtained with the SSCHA, compared to the extremely high \tcdw\ found at harmonic level (which, on the contrary, included only the electronic fluctuations), suggests that the ionic fluctuations play the dominant role in suppressing the CDW order. However, since in bulk and monolayer NbSe${}_2$ the electronic density of states at the Fermi level is quite  sizeable~\cite{C3RA47237J,PhysRevB.80.241108}, mainly due to localized $d$-states, the electronic contribution to entropy is not guaranteed to be negligible when anharmonicty is taken into account. In order to carefully estimate the electronic contribution to the fluctuations also at the anharmonic level, so as to give a quantitative comparison between the ionic and electronic thermal effects, we perform SSCHA calculations including finite-temperature electronic fluctuations. We consider the Hamiltonian $K+\Fel(\bR)$ for the ionic system (if electrons have finite temperature, in the adiabatic approximation forces and equilibrium position of the ions are ruled by the electronic free energy), and we minimize the functional
\begin{eqnarray}
F[\rhoscha] &=& \Avg{K+\Fel(\bR)}-T\,\Sion[\rhoscha]\, \nonumber \\ &=&\Avg{K+\Eel(\bR)}-T\,S[\rhoscha]\,,
\end{eqnarray}
where $S[\rhoscha]=\Avg{\Sel(\bR)}+\Sion[\rhoscha]\,$. Therefore, the SSCHA estimation of the system's entropy now includes contributions from both electrons (averaged through the ionic distribution) and ions. For the monolayer at 200~K, the difference between the SSCHA phonon frequencies obtained with and without the finite-temperature electronic contribution is around $2.6\%$, which has a negligible impact on the estimation of \tcdw. At lower (electronic) temperatures, since the harmonic results remain identical (see Supplementary Material\cite{supp}), the corrections to the SSCHA frequencies coming from the inclusion of finite-temperature electronic effects are not expected to be larger than that. Therefore, this test confirms that the ionic fluctuations totally overshadow the electronic fluctuations when anharmonicity is taken into account.

In conclusion, making use of first-principles calculations within the stochastic self-consistent harmonic approximation, we show that anharmonicity melts the CDW in both bulk and monolayer NbSe$_2$, and that ionic thermal fluctuations predominate over electronic thermal effects. Moreover, we show that, in spite of interlayer interaction, the CDW transition temperature and spatial modulation weakly depend on thickness.
%This situation, together with the fact that the same in-plane lattice modulation is found %in both bulk and monolayer experiments, which is in agreement with our calculations, %indicates the genuine 2D character of the CDW instability in NbSe$_2$.
%This is in contrast with what happens in other layered systems like, for example, %TiSe$_2$~\cite{Kolekar_2017} and VSe$_2$~\cite{PhysRevLett.121.196402}, where the %out-of-plane nature of the bulk CDW is related to a different \tcdw\ observed in the %corresponding monolayer. 
Our result supports the STM measurements on single layers grown by MBE~\cite{Ugeda2016}, but questions the Raman results on exfoliated samples that estimate a large enhancement of \tcdw\ in the monolayer~\cite{Xi2015,lin2020patterns}. It is to be seen whether the enhanced \tcdw\ observed in Raman experiments is a consequence of the sample deterioration, e.g., oxidation, during the exfoliation process or it is affected by the substrate. Indeed, similar theoretical calculations as those presented here have recently shown that strain or charge doping from the substrate can affect the CDW transition in other TMDs~\cite{doi:10.1021/acs.nanolett.9b00504,doi:10.1021/acs.nanolett.0c00597}.        

%
% Acknowledgements   
%

\vspace{0.5cm}

R. B. would like to thank Andrea Floris for fruitful discussions. I.E. acknowledges financial support from the Spanish Ministry of Economy and Competitiveness (Grant FIS2016-76617-P). M. C. acknowledges suppport from Agence Nationale de la Recherche, Grant N. ANR-19-CE24-0028. L.M., M. C. and F. M. acknowledge support from the Graphene
Flagship. F.M. and L.M. acknowledge  support by the MIUR PRIN-2017 program, project number 2017Z8TS5B.
Calculations were performed at the Joliot Curie-AMD supercomputer under the PRACE project RA4956 and at the supercomputers of the Donostia International Physics Center (DIPC). 

%\bibliography{bibliography}
%merlin.mbs apsrev4-1.bst 2010-07-25 4.21a (PWD, AO, DPC) hacked
%Control: key (0)
%Control: author (72) initials jnrlst
%Control: editor formatted (1) identically to author
%Control: production of article title (-1) disabled
%Control: page (0) single
%Control: year (1) truncated
%Control: production of eprint (0) enabled
%

\end{document}

%\title{Weak Dimensionality Dependence of the Charge Density Wave Transition in NbSe$_2$}
\title{Supplementary Material: \\ Weak Dimensionality Dependence and Predominant Role of Ionic Fluctuations in the Charge-Density-Wave Transition of NbSe$_2$}

\author{Raffaello Bianco$^{1}$}
\author{Lorenzo Monacelli$^{2,3}$}
\author{Matteo Calandra$^{3,4,5}$}
\author{Francesco Mauri$^{2,3}$}
\author{Ion Errea$^{1,6,7}$}

\affiliation{$^1$Centro de F\'isica de Materiales (CSIC-UPV/EHU), Manuel de Lardizabal pasealekua 5, 20018 Donostia/San Sebasti\'an, Spain}
\affiliation{$^2$Dipartimento di Fisica, Università di Roma Sapienza, Piazzale Aldo Moro 5, I-00185 Roma, Italy}
\affiliation{$^3$Graphene Labs, Fondazione Istituto Italiano di Tecnologia, Via Morego, I-16163 Genova, Italy}
\affiliation{$^4$ Dipartimento di Fisica, Università di Trento, Via Sommarive 14, 38123 Povo, Italy.}
\affiliation{$^5$ Sorbonne Universit\'e, CNRS, Institut des Nanosciences de Paris, UMR7588, F-75252 Paris, France}
\affiliation{$^6$Fisika Aplikatua 1 Saila, Gipuzkoako Ingeniaritza Eskola, University of the Basque Country (UPV/EHU),
Europa Plaza 1, 20018 Donostia/San Sebasti\'an, Spain}
\affiliation{$^7$Donostia International Physics Center (DIPC), Manuel Lardizabal pasealekua 4, 20018 Donostia/San Sebasti\'an, Spain}

\maketitle

\renewcommand{\figurename}{SUPPLEMENTARY FIG.}

\section{The stochastic Self-consistent harmonic approximation}

The stochastic self-consistent harmonic approximation (SSCHA)~\cite{PhysRevB.89.064302,PhysRevB.96.014111,PhysRevB.98.024106} is a quantum variational method that minimizes the free energy of the system $F$ with respect to centroid positions $\rscha$ and effective force constants $\phischa$ as described in the main text. In the static limit of the SSCHA~\cite{PhysRevB.96.014111} the phonon frequencies are obtained diagonalizing the Hessian of the SSCHA free energy $F$, $\frac{1}{\sqrt{M_aM_b}} \frac{\partial^2 F}{\partial \mathcal{R}_a \partial \mathcal{R}_b}$, where $a$ and $b$ label both an atom and a Cartesian index and $M_a$ is the mass of atom $a$. The free energy Hessian in the SSCHA is explicitly given in compact notation as~\cite{PhysRevB.96.014111}
\begin{equation}
    \frac{\partial^2 F}{\partial \rscha \partial \rscha} = \phischa + \overset{(3)}{\phischa} \boldsymbol{\Lambda}\bigg[\mathbf{1} - \boldsymbol{\Lambda}\overset{(4)}{\phischa}\bigg]^{-1} \overset{(3)}{\phischa}.
    \label{sscha-hessian}
\end{equation}
In Eq. \eqref{sscha-hessian}  $\Phi_{ab}$ are the SSCHA variational force constants. $\Lambda_{abcd}$ is a fourth order tensor that depends on temperature as well as on the eigenvectors and eigenvalues of $\phischa$. The $ \overset{(n)}{\phischa}$ tensors are defined as the quantum statistical averages of the $n$-th order derivatives of the Born-Oppenheimer potential $V$ taken with respect to the SSCHA density matrix $\rho_{\rscha,\phischa}$:
\begin{equation}
    \overset{(n)}{\Phi}_{a_1 \dots a_n} = \left\langle \frac{\partial^n V}{\partial R_{a_1} \dots \partial R_{a_n} }\right\rangle_{\rho_{\rscha,\phischa}}.
    \label{phin}
\end{equation}

As mentioned in the main text, the application of the SSCHA requires the calculation of atomic forces on supercells. For a supercell of size $n\times m \times l$, the SSCHA will include the phonon-phonon interaction of all the modes on a commensurate $n\times m \times l$ grid of points in the Brillouin zone.

\section{Details of the {\it ab initio} calculations}

The force calculations on supercells needed to carry out the SSCHA minimizations both in bulk and monolayer NbSe$_2$ were performed within density functional theory (DFT), assuming the Perdew-Burke-Ernzerhof (PBE) parametrization of the exchange-correlation functional~\cite{PhysRevLett.77.3865}, and employing the {\sc Quantum ESPRESSO} package~\cite{0953-8984-21-39-395502,0953-8984-29-46-465901}. All SSCHA results presented in the main text were obtained with a $6\times 6 \times 1$ supercell both for the bulk and the monolayer. Calculations with other supercells were also performed (see below). DFT calculations were performed with an ultrasoft pseudopotential for Nb  and a norm conserving 
pseudopotential for Se, with a 35 Ry energy cutoff for the plane-wave basis, and a 400 Ry cutoff for the density. Harmonic phonon calculations were carried out within density-functional perturbation theory (DFPT)~\cite{RevModPhys.73.515}. 

\begin{figure}[t]
\includegraphics[width=1.0\columnwidth]{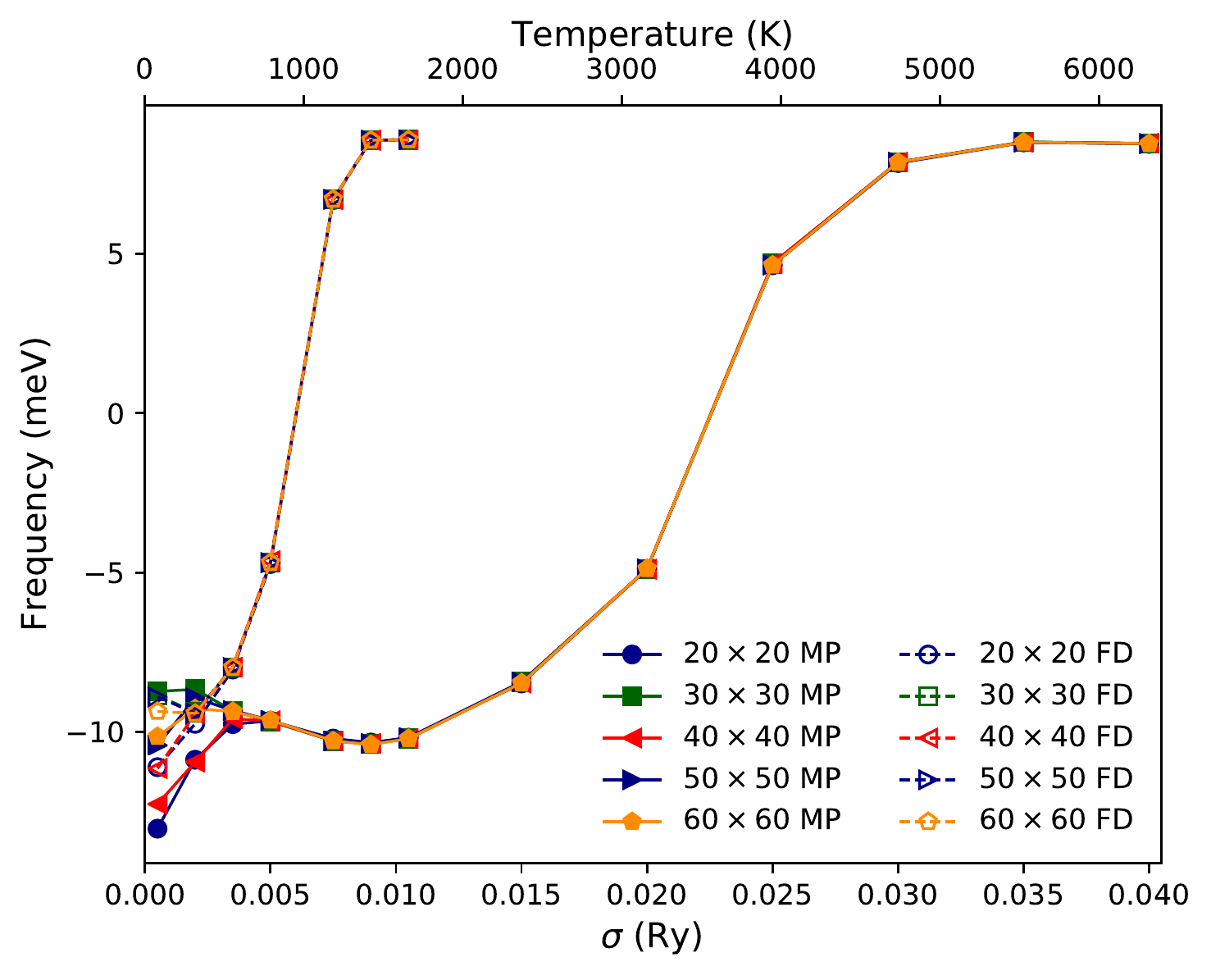}
\caption{Harmonic phonon frequency of the longitudinal acoustic mode at \qcdw$=2/3\Gamma M$ for the NbSe$_2$ monolayer that drives the CDW transition as a function of the Methfessel-Paxton smearing parameter $\sigma$ with different grids for the Brillouin zone integrals. We also show the same results calculated with the Fermi-Dirac distribution function at different temperatures.} 
\label{figSI1}
\end{figure}

The integration over the Brillouin zone for the DFPT hamonic calculation in the bulk was performed with a $24 \times 24 \times 8$ grid and a Methfessel-Paxton smearing of 0.01 Ry. These parameters were shown to be converged in a previous work~\cite{PhysRevB.92.140303}. For the monolayer, we show in Supplementary Fig. \ref{figSI1} that a finer $40 \times 40$ grid and a smaller value of 0.005 Ry for the smearing are required to converge the phonon frequency of the longitudinal acoustic mode at \qcdw$=2/3\Gamma M$ that drives the CDW transition. In all these calculations the electronic systems is assumed to be at 0 K. Harmonic phonon calculations with the converged grid but with the explicit Fermi-Dirac distribution were also performed as shown in Figs. 2 and 4 of the main text, in order to study the effect of the electronic entropy on the harmonic phonons. In Supplementary Fig. \ref{figSI1} we also show the dependence on temperature of the harmonic phonons calculated with the explicit Fermi-Dirac distribution for different integration grids. Below approximately 400 K the harmonic frequency of the CDW mode saturates: it does not change anymore for the largest integration grids and matches the converged value obtained with Methfessel-Paxton smearing technique. 

The Brillouin zone integration for the force calculations in the distorted supercells required coarser meshes to converge the SSCHA minimizations. For the bulk, grids that would correspond to a $12 \times 12 \times 4$ grid in the unit cell (e.g. a $2 \times 2 \times 4$ grid for a $6 \times 6 \times 1$ supercell) and a Methfessel-Paxton smearing of 0.01 Ry are enough. For the monolayer, grids that would correspond to a $30 \times 30 \times 1$ grid in the unit cell (e.g. a $5 \times 5 \times 1$ grid for a $6 \times 6 \times 1$ supercell) and a Methfessel-Paxton smearing of 0.005 Ry are used throughout.     

In order to consider both the electronic and the ionic entropy in the anharmonic SSCHA calculations, we also performed SSCHA minimizations with DFT forces calculated considering the electronic entropy by including the finite-temperature Fermi-Dirac occupation. We limited these calculations to the $6\times 6$ monolayer supercell at 200 K. As mentioned in the main text, including the electronic entropy in the calculations barely affects the SSCHA result, showing that once ionic fluctuations are included the electronic entropy is completely overshadowed in the temperature region of interest.

The calculations are performed with the experimental lattice parameter for the bulk~\cite{MAREZIO1972425}: $a=3.44\AA$ and $c=12.55\AA$. For the monolayer we use the same $a=3.44\AA$ and an interlayer distance of $c=12.55\AA$ (note that one layer is missing here). We did not consider thermal expansion. Symmetry was preserved in the SSCHA minimizations. The internal coordinates were relaxed classically to the minimum of the Born-Oppenheimer potential and were kept fixed in the SSCHA minimizations. We checked that the SSCHA barely affects internal coordinates, and assuming them fixed does not alter the calculated \tcdw. 

\section{Convergence of the Fourier interpolated harmonic phonon spectra}

\begin{figure}[t]
\includegraphics[width=1.0\columnwidth]{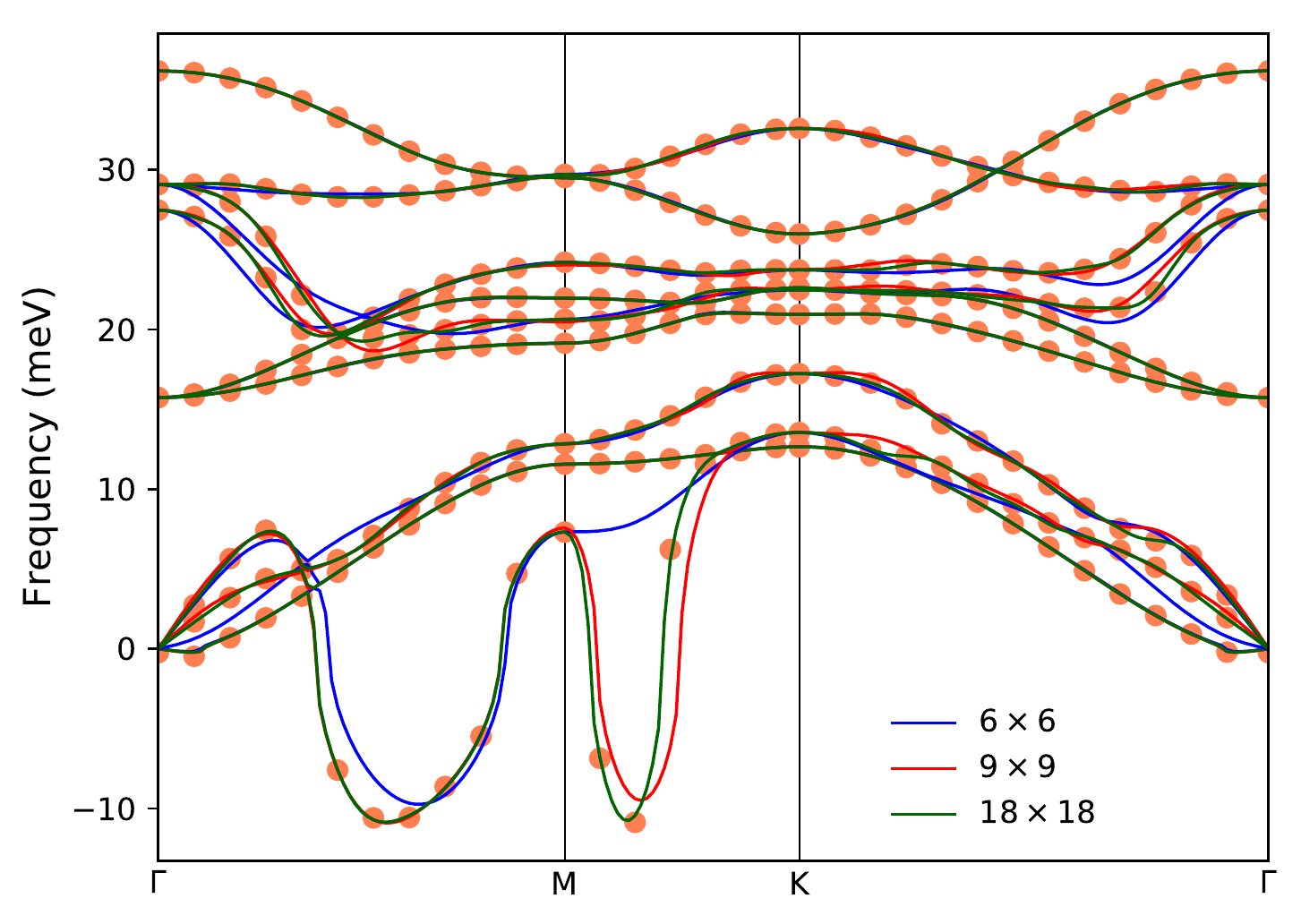}
\caption{Harmonic phonon spectra for the NbSe$_2$ monolayer obtained with Fourier interpolation making use of different grids. The results are compared with a point by point DFPT calculation (orange dots).} 
\label{figSI2}
\end{figure}

In Supplementary Fig. \ref{figSI2} we show the harmonic phonon spectra obtained for the monolayer by Fourier interpolating the force constants obtained in $6\times 6$, $9 \times 9$, and $18 \times 18$ grids and calculated point by point by DFPT. The results show that converging the Fourier interpolated spectra requires a very large interpolation grid. While the $18 \times 18$ grid is fully converged as it matches the point by point calculation, the $6\times 6$ grid misses the instability along $MK$. The $9 \times 9$ grid does not provide a fully converged spectra close to the harmonic instabilities. 

\section{Dependence of the harmonic phonon spectra on the smearing parameter}

\begin{figure}[t]
\includegraphics[width=1.0\columnwidth]{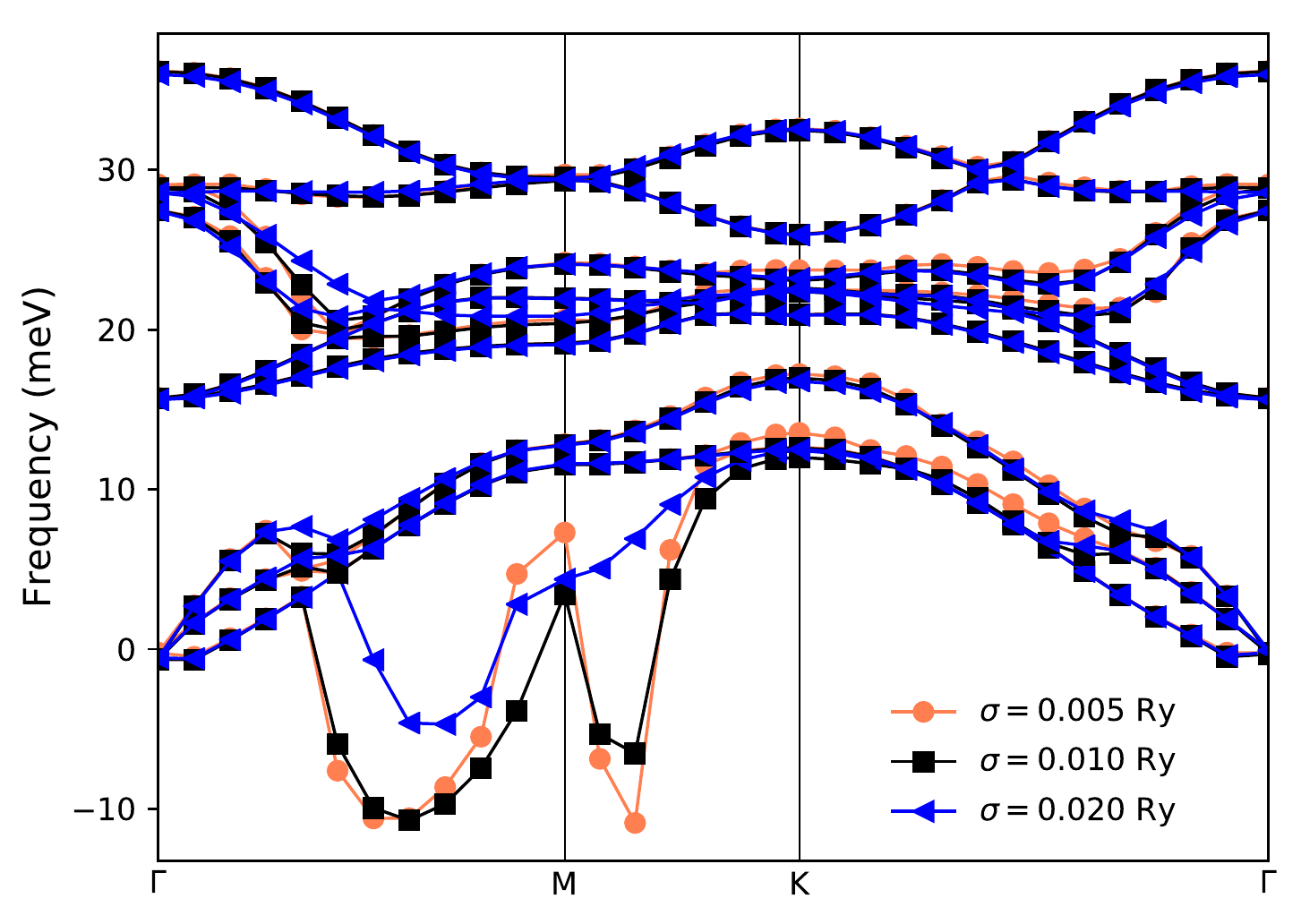}
\caption{Harmonic phonon spectra for the NbSe$_2$ monolayer calculated point by point using a $40 \times 40$ grid and different Methfessel-Paxton smearing parameters $\sigma$ for the Brillouin zone integrals.} 
\label{figSI3}
\end{figure}

In Supplementary Fig. \ref{figSI3} we show that the harmonic phonon spectra strongly depends on the  Methfessel-Paxton smearing parameter used for the Brillouin zone integrals. It is clearly seen that increasing the smearing parameter the instability along $MK$ is washed out and the instability along $\Gamma M$ is shifted to a larger momentum, as it happens when increasing the temperature in the harmonic calculation with the Fermi-Dirac distribution function. However, as shown in Supplementary Fig. \ref{figSI1}, these are not converged results in the 0 K limit.

\section{Analysis of the Fourier interpolated anharmonic SSCHA phonon spectra}

\begin{figure}[t]
\includegraphics[width=1.0\columnwidth]{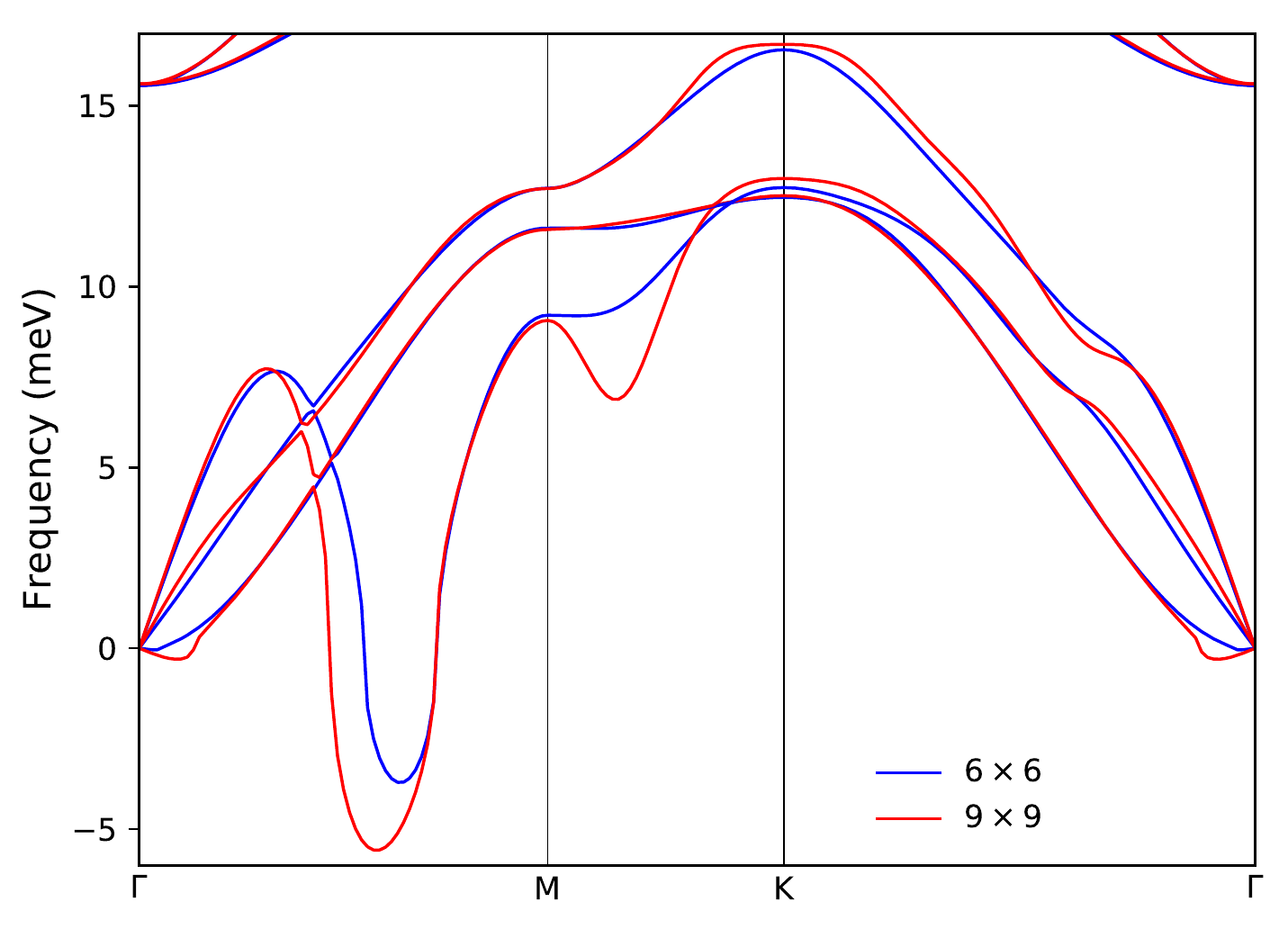}
\caption{Anharmonic phonon spectra at 50 K for the NbSe$_2$ monolayer obtained by Fourier interpolating the SSCHA free energy Hessian force constants obtained with $6 \times 6$ and $9 \times 9$ supercells.} 
\label{figSI4}
\end{figure}

In Supplementary Fig. \ref{figSI4} the phonon spectra of the monolayer obtained by Fourier interpolating the Hessian of the SSCHA free energy (see Eq. \eqref{sscha-hessian}) obtained with different supercells. The results show that even if at the $\mathbf{q}$ points commensurate with the supercell the phonon spectra is converged, the Fourier interpolation yields different results at the instability along $\Gamma M$, i.e., the Fourier interpolated spectrum is not converged with the supercell size yet. 

The calculation in the $9 \times 9$ supercell shows clearly that the instability along $MK$, which can be captured by this grid (see Suppelementay Fig. \ref{figSI2}), is completely washed out by anharmonicity. The $6 \times 6$ supercell calculation is not enough to determine this, as it misses this instability (see Suppelementay Fig. \ref{figSI2}).    
 
\section{Supercell dependence of \tcdw and the effect of the different terms on the free energy Hessian}

\begin{figure}[t]
\includegraphics[width=1.0\columnwidth]{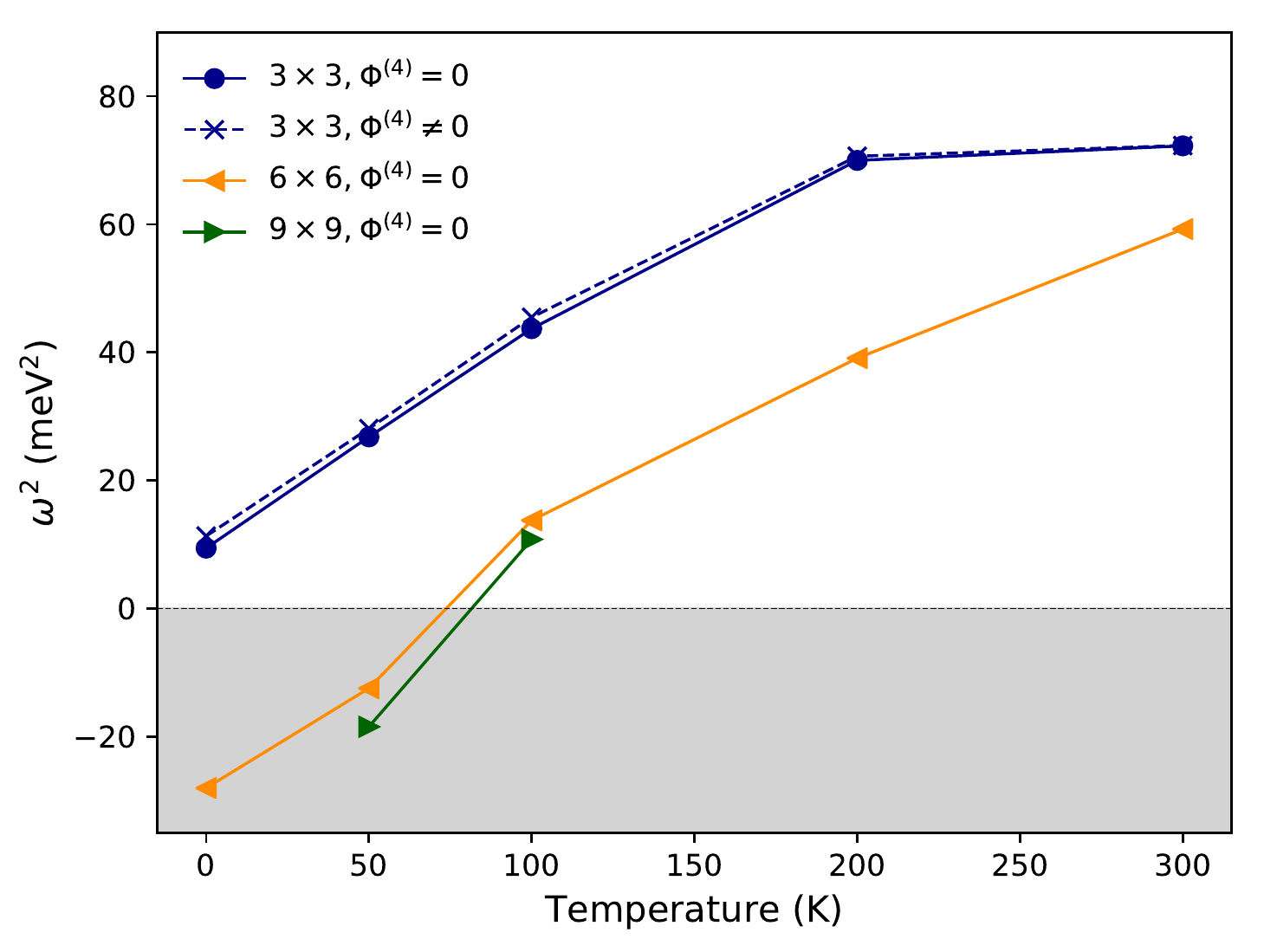}
\caption{Anharmonic phonon frequency of the longitudinal acoustic mode at \qcdw$=2/3\Gamma M$ for the NbSe$_2$ monolayer obtained from the SSCHA free energy Hessian calculated with different supercells. For the $3 \times 3$ case, the effect of including $\overset{(4)}{\phischa}$ in Eq. \eqref{sscha-hessian} is explicitly shown.} 
\label{figSI5}
\end{figure}

Even if the anharmonic Fourier interpolated phonon spectrum is not fully converged, the phonon frequency of the LA mode at \qcdw$=2/3\Gamma M$ is fully converged with the $6 \times 6$ supercell. In Supplementary Fig. \ref{figSI5} we show its phonon frequency as a function of temperature as calculated with $3 \times 3$, $6 \times 6$, and $9 \times 9$ supercells for the monolayer. This \qcdw\ point is commensurate with all these supercells. The results clearly show that, while the $3 \times 3$ supercell result is far from convergence, the $6 \times 6$ supercell is enough to determine \tcdw\ from the frequency of longitudinal acoustic mode at \qcdw. The \tcdw\ value obtained with the $9 \times 9$ supercell is 81 K, only 8 K more than the $6 \times 6$ supercell result. 

All anharmonic phonon spectra calculated from the Hessian of the SSCHA free energy in this work have been obtained setting $\overset{(4)}{\phischa}=0$ in Eq. \eqref{sscha-hessian}. In Supplementary Fig. \ref{figSI5} it can be seen that including $\overset{(4)}{\phischa}$ in the equation has a negligible effect on \tcdw\ for the $3 \times 3$ supercell in the monolayer case. A similar behavior is expected for the bulk and other supercells.

Finally, in Supplementary Fig. \ref{figSI6} we compare for the monolayer the harmonic phonon spectra with the anharmonic phonons obtained with the free energy Hessian (with $\overset{(4)}{\phischa}=0$) and just with the SSCHA auxiliary force constants $\phischa$ (which would imply making also $\overset{(3)}{\phischa}=0$ in Eq. \eqref{sscha-hessian}). The spectra coming from $\phischa$ is positive definite by definition. From the figure it is clear that the anharmonic correction as well as the temperature dependence of the phonons concentrates in the low-energy region.  

\begin{figure}[t]
\includegraphics[width=1.0\columnwidth]{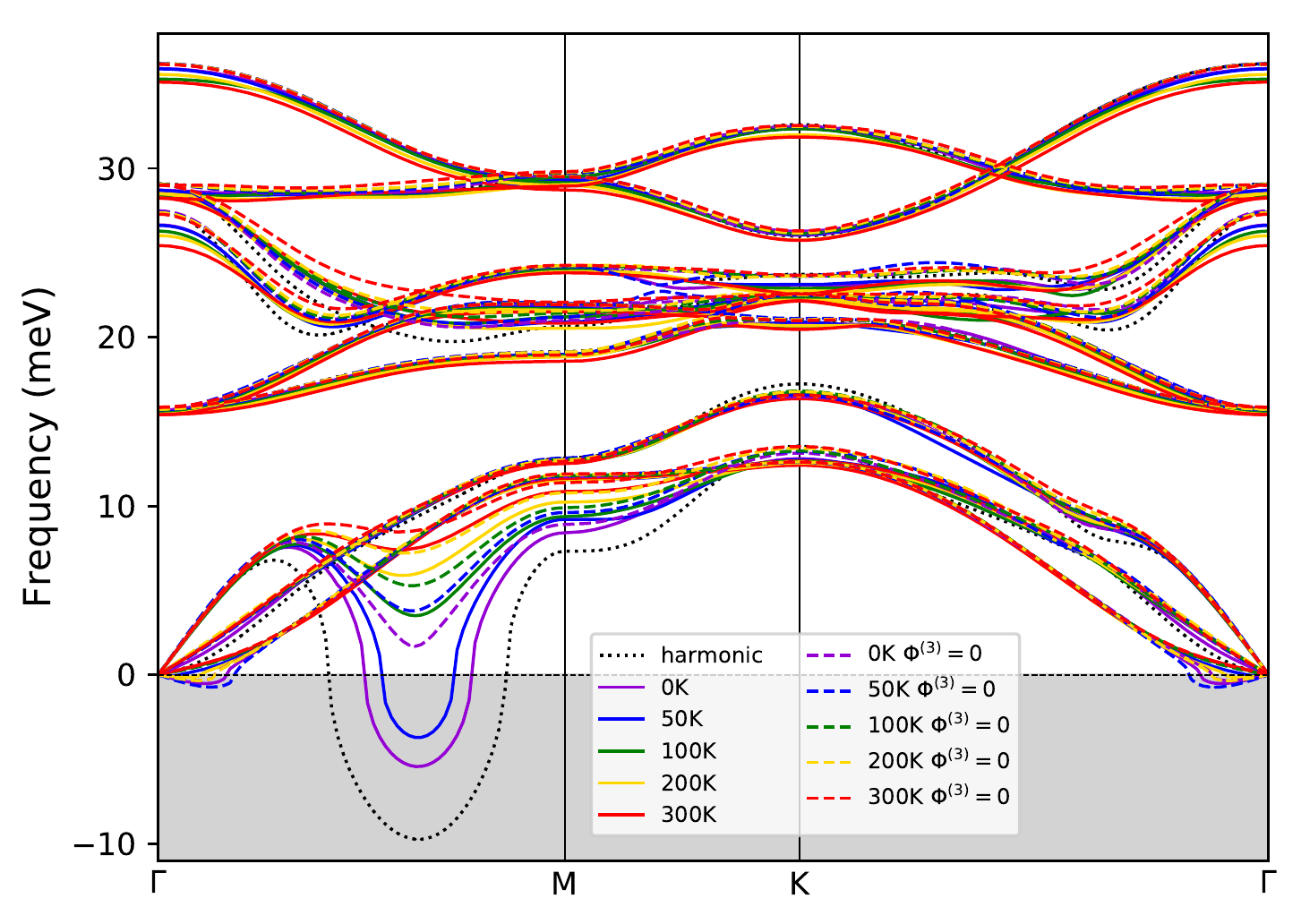}
\caption{Full harmonic phonon spectra together with the anharmonic phonon spectra at different temperatures for the monolayer of NbSe$_2$. The anharmonic results are obtained directly from the SCHA auxiliary force constants $\phischa$, which implies setting $\overset{(3)}{\phischa}=0$ in Eq. \eqref{sscha-hessian} (labelled accordingly), and from the free energy Hessian (already presented in the main text).} 
\label{figSI6}
\end{figure}

\section{Phonon distortion pattern in the bulk at \qcdw\ for the CDW mode and the other low-energy mode}

In Supplementary Fig. \ref{figSI7} we show the distortion patterns associated to the two lowest energy modes at \qcdw$=2/3\Gamma M$ for the bulk. In the harmonic approximation these two modes are unstable at low temperature. The lowest energy mode is the CDW mode, while the other mode is never imaginary in the anharmonic SSCHA calculations. Both modes are commensurate with the $3 \times 3 \times 1$ supercell pattern shown in the figure. The chosen distortion patterns correspond to one of the degenerate modes associated to all the $\mathbf{q}$ points related by symmetry to \qcdw. Both distortion patterns are similar but the distortions in each layer are not identical for the two modes. In the absence of interlayer interactions these two modes should be degenerate in energy, but this is not the case, showing the presence of interlayer interactions. 

\begin{figure}[t!]
\includegraphics[width=1.0\columnwidth]{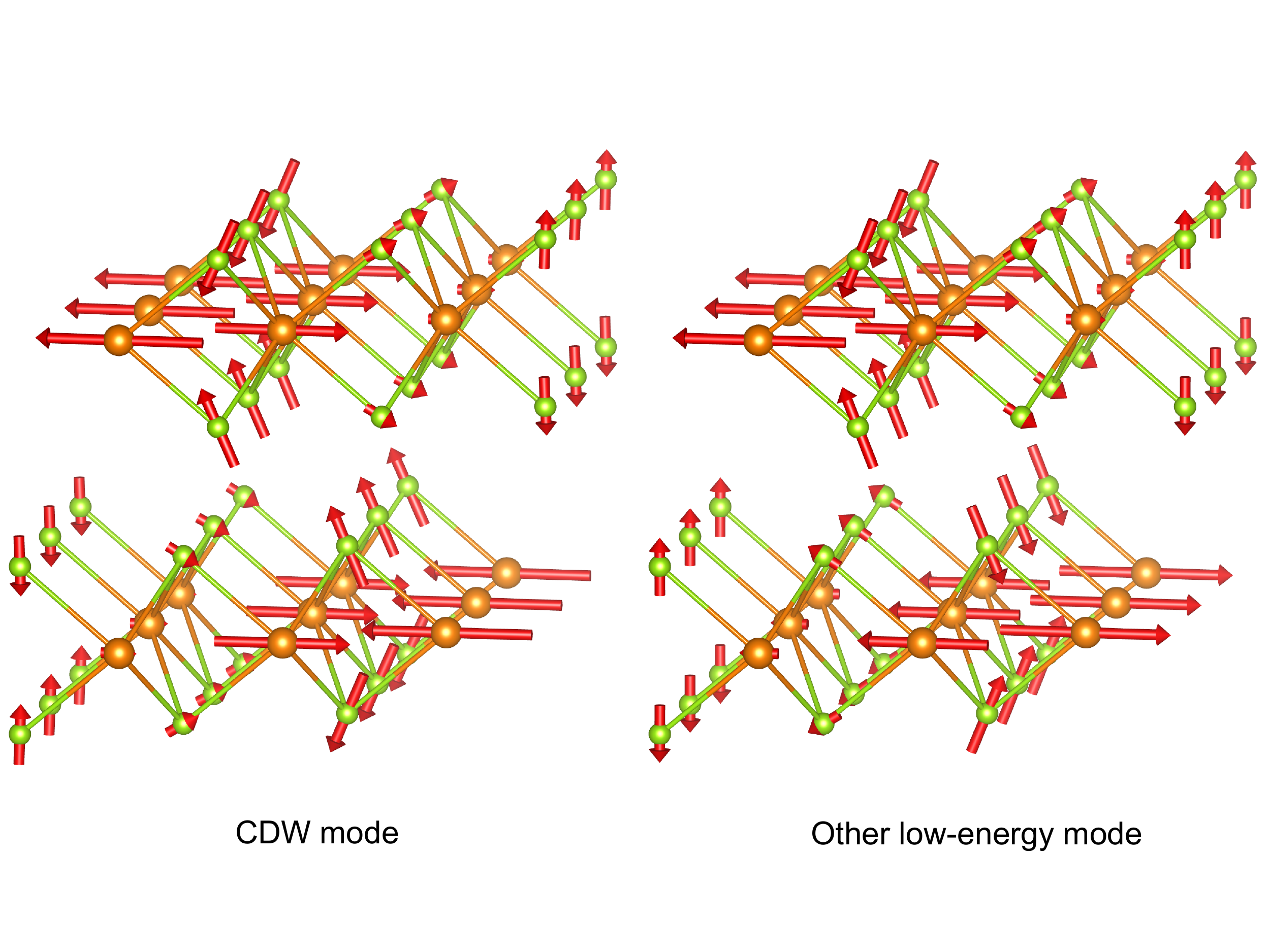}
\caption{Real space distortion patterns associated to the two lowest energy modes at \qcdw$=2/3\Gamma M$ for the bulk.} 
\label{figSI7}
\end{figure}

\bibliography{bibliography}